\documentclass[letterpaper]{jpconf}
\pdfoutput=1

\usepackage{graphicx}
\usepackage{exscale}
\usepackage{color}

\begin{document}
\title{Searches for New Physics Using H1 and ZEUS Data}

\author{Antje H\"uttmann}

\address{Deutsches Elektronen-Synchrotron DESY}
\address{on behalf of the H1 and ZEUS collaborations}

\begin{abstract}
Recent searches for new physics in $ep$ collisions performed by the H1 and ZEUS collaborations are presented. Limits on different contact interaction models, large extra dimensions, $R$-parity violating SUSY, excited fermions and anomalous flavour-changing top couplings are shown. In addition, searches for new physics in exclusive final states such as events with isolated leptons and large missing transverse momentum or multi-lepton final states are presented. 
\end{abstract}

\section{Introduction}
At the HERA collider electrons or positrons with a beam energy of 27.5~GeV/c and protons with a beam energy of 920~GeV/c were brought into collision, resulting in a centre-of-mass energy of 319~GeV/c. HERA was in operation from 1992 to 2007 and up to now was the only electron-proton collider in the world. Thus it offers unique opportunities to search for physics beyond the Standard Model (SM). The two collider experiments ZEUS and H1 each collected 0.5~$\mathrm{fb}^{-1}$ of data and are combining this data to improve the precision of the measurements. 

\section{Contact Interactions}
\subsection{General Model}
Contact interactions are an effective theory describing low energy effects from physics at much higher energy scales, $\Lambda \gg \sqrt{s}$. They could alter the SM Deep Inelastic Scattering (DIS) distributions at high squared four-momentum transfer $Q^2$. Vector-type $eeqq$ contact interactions are described by an effective Lagrangian:
\begin{equation}
\mathcal{L}_{CI} = \sum_{a,b=L,R}^{q=u,d} \eta_{ab}^q (\overline{e}_a \gamma_\mu e_a) (\overline{q}_b \gamma^\mu q_b),
\end{equation}
where the sums run over the polarisations of electrons and quarks ($L$ meaning left-handed and $R$ meaning right-handed polarisation) and the quark flavour (up quarks or down quarks) and $\eta_{ab}^q$ is given by  
\begin{equation}
\eta_{ab}^q = \pm 4\pi / \Lambda^2.
\end{equation}
Many variants of the model can be formulated, which differ by their chiral structure. Since no deviations from the SM prediction for NC DIS were observed, ZEUS set lower limits on $\Lambda$ for 19 models with different helicity structure, reaching from 3.8~TeV/c to 8.9~TeV/c, at 95\%~CL \cite{CI}. 

\subsection{Large Extra Dimensions}
In the ADD (Arkani-Hamed, Dimopoulos, Dvali) model \cite{ArkaniHamed:1998rs}, the space-time is $4+n$ dimensional. SM particles are confined to four dimensions, while gravity can propagate into the extra dimensions. The fundamental Planck scale $M_S$ in $4+n$ dimensions can be of the order of 1~TeV/c. In this case the strength of gravitational and electroweak interactions would be comparable at high energies and the hierarchy problem would be solved. A virtual graviton exchange contribution to $eq \rightarrow eq$ scattering can be described by a contact interaction with effective coupling $\eta_G \sim \pm 1 / M_S$. ZEUS set a lower limit on $M_S$ of 0.94~TeV/c at 95\%~CL \cite{CI}, for both signs of $\eta_G$. Fig. \ref{fig:squarks} (left) shows the ZEUS $e^+p$ data compared to the exclusion limits.

\section{Squark Production in $R$-parity violating supersymmetry}
\begin{figure}[hbt]
\begin{center}
\includegraphics[width=8cm]{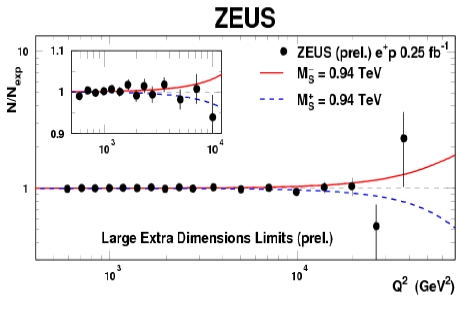}
\includegraphics[width=7.5cm]{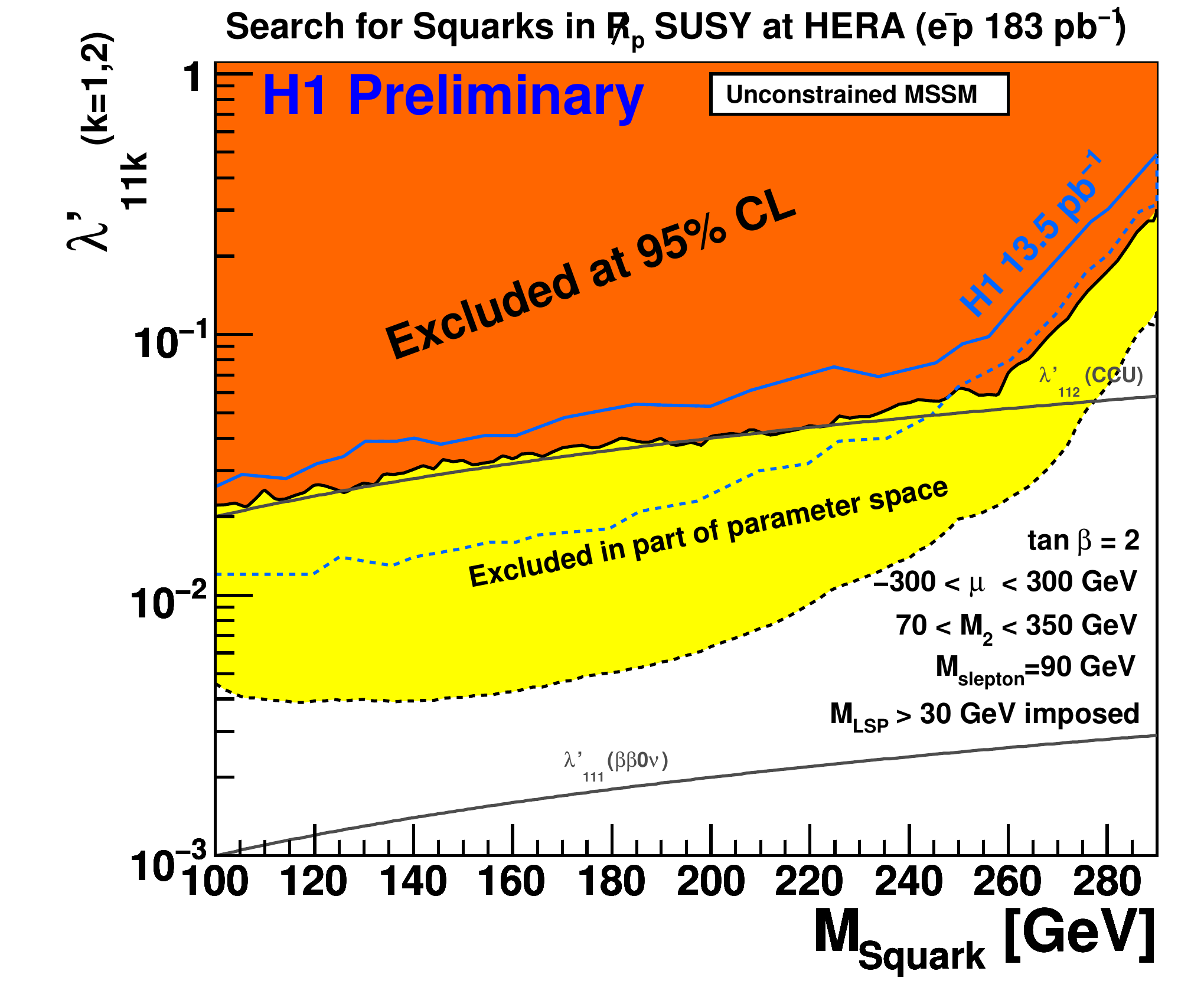}
\caption{ZEUS $e^+p$ data compared to the 95\% CL exclusion limits on the Planck mass scale in models with large extra dimensions, for positive ($M_S^+$) and negative ($M_S^-$) couplings (left plot), and the exclusion limit on the Yukawa coupling $\lambda^{\prime}$ as a function of the $\tilde{d}_R$, $\tilde{s}_R$ squark mass (right plot).}
\label{fig:squarks}
\end{center}
\end{figure}

In SUSY models with $R$-parity violation single squarks can be produced resonantly in $ep$ collisions. The squarks decay either to lepton and quark giving DIS-like final states or to states involving a gaugino, leading to cascade decays. Several final state topologies corresponding to about 90\% of the branching ratio were analyzed and no deviations from the SM were observed. Thus limits were set using the full H1 data \cite{Squarks}. For first and second generation squarks and a Yukawa coupling of the electromagnetic strength, $\tilde{u}_L$, $\tilde{c}_L$ ($\tilde{d}_R$, $\tilde{s}_R$) squarks with masses up to $\sim275~\mathrm{GeV/c^2}$ ($\sim290~\mathrm{GeV/c^2}$) are excluded at 95\% CL. Fig. \ref{fig:squarks} (right) shows the exclusion limit on the Yukawa coupling $\lambda^{\prime}$ as a function of the $\tilde{d}_R$, $\tilde{s}_R$ squark mass. 

\section{Excited Fermions} 
\begin{figure}[hbt]
\begin{center}
\includegraphics[width=7cm]{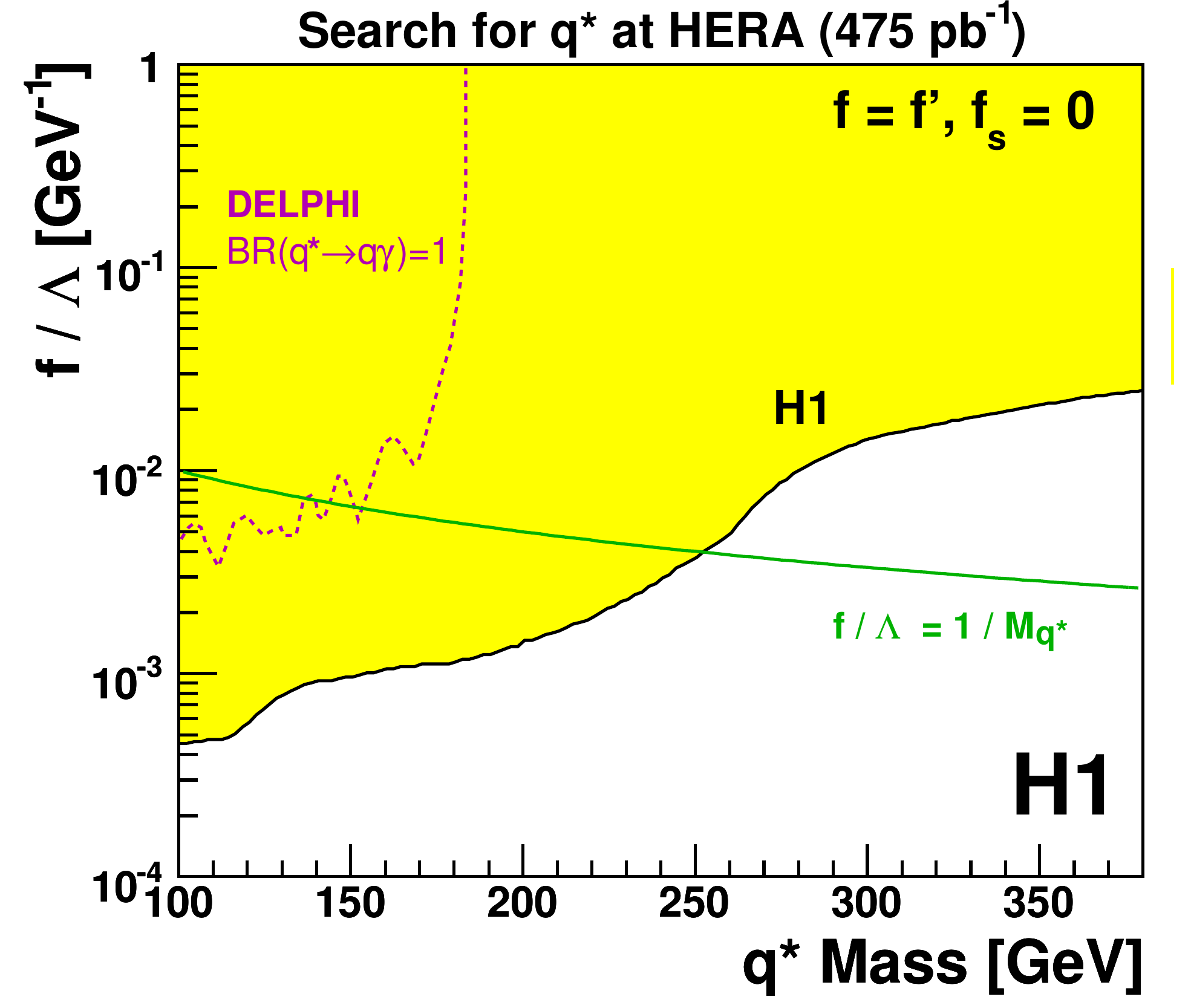}
\caption{Exclusion limits on the ratio of the coupling to the compositeness scale $f/\Lambda$ as a function of the $q^*$ mass at 95\% CL.}
\label{fig:excited}
\end{center}
\end{figure}

The observation of excited fermions would be direct evidence for compositeness (i.e. fermion substructure). Compositeness could explain the three lepton and the three quark families as well as their mass hierarchy. A search for excited electrons, neutrinos and quarks was performed by the H1 collaboration. Excited fermions would decay to standard fermions and gauge bosons, therefore subsequent leptonic and hadronic decay channels of $Z$ and $W$ bosons were investigated. No evidence for excited fermion production was found and thus limits were set \cite{:2008xe,Aaron:2008cy,Aaron:2009iz} on the ratio of the coupling to the compositeness scale, $f/\Lambda$, at 95\% CL as a function of the mass of the excited fermion. The limit for excited quarks is shown in Fig. \ref{fig:excited}. Assuming $f/\Lambda = 1/M_{f^*}$, excited quarks (electrons, neutrinos) with masses up to $252~\mathrm{GeV/c^2}$ ($272~\mathrm{GeV/c^2}$, $213~\mathrm{GeV/c^2}$) are excluded at 95\% CL. 

\section{Isolated Leptons and Missing Transverse Momentum\label{sec:IsoLep}} 
Events with isolated leptons and large missing transverse momentum in the final state are rare but have a clean signature and thus are sensitive to new physics. The dominating SM process with this final state topology is single-$W$ production. Using the combined H1 and ZEUS data \cite{:2009wg}, a few interesting events at high transverse momentum of the hadronic system, $p_T^X > 25$~GeV/c, were observed mostly in the $e^+p$ data (23 events compared to $14.0 \pm 1.9$ events expected within the SM). 

\section{Anomalous Single Top Production} 
\begin{figure}[hbt]
\begin{center}
\includegraphics[width=7cm]{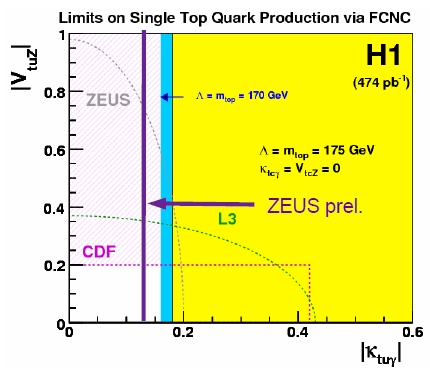}
\caption{Excluded regions of the anomalous coupling $\kappa_{tu \gamma}$-$v_{tuZ}$ plane.}
\label{fig:ST}
\end{center}
\end{figure}

At HERA top quarks can only be singly produced. The SM cross section is negligible, but the production of single top quarks via an anomalous Flavour-Changing-Neutral-Current coupling is predicted by several beyond the Standard Model theories. Thus the observation of top quarks at HERA would be a clear indication of new physics. A leptonic top decay ($t \rightarrow bW^+ \rightarrow be^+\nu$) would lead to final states with an isolated lepton, large missing transverse momentum and large hadronic transverse momentum as mentioned in Section \ref{sec:IsoLep}. Since no significant deviations from the SM in the leptonic or the hadronic decay channels are seen by H1 \cite{Aaron:2009vv} and ZEUS \cite{Chekanov:2003yt,Chekanov:2003zh,ST}, limits were set on the anomalous couplings $\kappa_{tu \gamma}$ and $v_{tuZ}$ (Fig. \ref{fig:ST}). The HERA limits on $\kappa_{tu \gamma}$ are the most stringent limits available. 

\section{Multi-Lepton Events\label{sec:MultiLep}} 
The main SM production processes for events with multi-lepton final states are photon-photon interactions with two or all three final state leptons detected. The SM expectation is small at high invariant mass and high $p_T$ of the leptons and thus new phenomena are expected to be visible in this region. Also in this case, using the combined H1 and ZEUS data \cite{:2009sma}, a few interesting events with $\sum p_T > 100$~GeV/c were observed in the $e^+p$ data (7 events compared to $1.94 \pm 0.17$ events expected within the SM). 

\section{General Searches}
A model-independent generic search for deviations from the SM in final-state configurations involving at least two high-$p_T$ particles (electrons, muons, jets, photons or neutrinos) was performed by the H1 collaboration \cite{:2008ww}. In 27 of the possible configurations events were observed. All multi-lepton events mentioned in Section \ref{sec:MultiLep} which are located in the phase space of this analysis are found. A search for deviations from the SM in the event yields and in the sum of the transverse momenta and the invariant mass of the final state particles is performed. A statistical analysis is used to quantify the significance of the deviations. Good agreement with the SM is found and all deviations are consistent with statistical fluctuations. Furthermore the number of fluctuations given the large number of search channels is consistent with the SM. 

\section{Conclusions}
Recent searches for new physics in $ep$ collisions performed by the H1 and ZEUS collaborations have been reported. In general good agreement with the SM prediction is observed for all the investigated phenomena, involving contact interactions, large extra dimensions, $R$-parity violating SUSY, excited fermions, anomalous single top production and searches for new physics in exclusive final states such as events with isolated leptons and large missing transverse momentum or multi-lepton final states. 

\section*{References}
\bibliography{cites}

\end{document}